\documentclass[aps,prb,twocolumn, showpacs,superscriptaddress]{revtex4}

\usepackage{graphicx} 
\usepackage{multirow}
\usepackage[T1]{fontenc}
\usepackage[latin9]{inputenc}
\usepackage{amsmath}
\usepackage{amssymb}
\usepackage{color}
\usepackage{rotating}

\begin{document}
\title{Adsorption of Alkali, Alkaline Earth and Transition Metal Atoms
on Silicene}

\author{H. Sahin}\email{hasan.sahin@ua.ac.be}
\affiliation{Department of Physics, University of Antwerp, Groenenborgerlaan
171, B-2020 Antwerp, Belgium}

\author{F. M. Peeters}\email{francois.peeters@ua.ac.be}
\affiliation{Department of Physics, University of Antwerp, Groenenborgerlaan
171, B-2020 Antwerp, Belgium}
\date{\today}

\begin{abstract}

The adsorption characteristics of alkali, alkaline earth and transition
metal adatoms on silicene, a graphene-like monolayer structure of silicon, are
analyzed by means of first-principles calculations. In contrast to graphene,
interaction between the metal atoms and the silicene surface is quite strong
due to its highly reactive buckled hexagonal structure. In addition to
structural properties, we also calculate the electronic band dispersion, net
magnetic moment, charge transfer, workfunction and dipole moment of the metal
adsorbed silicene sheets. Alkali metals, Li, Na and K, adsorb to hollow site
without any lattice distortion. As a consequence of the significant charge
transfer from alkalis to silicene metalization of silicene takes place. Trends
directly related to atomic size, adsorption height, workfunction and dipole
moment of the silicene/alkali adatom system are also revealed. We found that the
adsorption of alkaline earth metals on silicene are entirely different from
their adsorption on graphene. The adsorption of Be, Mg and Ca turns silicene
into a narrow gap semiconductor.
Adsorption characteristics of eight transition metals Ti, V, Cr, Mn, Fe, Co, Mo
and W are also investigated. As a result of their partially occupied $d$
orbital, transition metals show diverse structural, electronic and magnetic
properties. Upon the adsorption of transition metals, depending on the adatom
type and atomic radius, the system can exhibit metal, half-metal and
semiconducting behavior. For all metal adsorbates the direction of the charge
transfer is from adsorbate to silicene, because of its high surface reactivity.
Our results indicate that the reactive crystal structure of silicene provides a
rich playground for functionalization at nanoscale.

\end{abstract}
\maketitle

\section{Introduction}

Recent advances in controllable synthesis and characterization of nanoscale
materials, have opened up important possibilities for the investigation of
ultra-thin two-dimensional systems. Chiefly the research 
efforts directed towards graphene\cite{novo, geim} have dominated the new era of two-dimensional 
materials. Many exceptional features of atomically thin graphene layers such as massless Dirac fermions, 
strength of the lattice structure, high thermal conductivity and half-integer Hall conductance have been 
revealed so far.\cite{kats,novo2,zhang, berger} In spite of its unique properties, due to the lack of a 
band gap and its weak light adsorption, graphene research efforts have focused on graphene composites 
over the past five years. Studies have demonstrated the existence of several chemically converted 
graphene structures such as grapheneoxide (GO),\cite{dikin, eda, robinson}
graphane (CH),\cite{elias, sofo, apl2009, prb2010, graphane-m}
fluorographene (CF),\cite{nair, robinson2, withers, leenaerts,hasancf}
and chlorographene (CCl).\cite{cl-exp, hasan-cl, mari-cl} High-
quality insulating behavior, thermal stability and extraordinary mechanical strength of fluorographene 
(CF) has inspired intense research on halogenated graphene derivatives.

Unusual properties of graphene promising for a variety of novel
applications\cite{c-ads,coating,mesh,sefa1,sefa2,sefa3}
have also triggered significant interest in one or several atom-thick honeycomb
structures of binary compounds. Early experimental studies aiming to synthesize
and characterize novel monolayer materials have revealed that graphene-like
sheets of BN are also stable.\cite{bn-synthesis, bn-prl, bn-nature} Though BN
has the same planar structure as graphene due to the ionic character of B-N
bonds, BN crystal is a wide band gap insulator with an energy gap of 4.6
eV.\cite{Guo,louie-nanoletter,barone2,bn-mehmet} The perfect lattice matching
between graphene and BN layers make it possible to construct nanoscale devices.
\cite{novoselov-nobel} Following the synthesis of hexagonal monolayer
of ZnO,\cite{znobilayer} the II-VI metal-oxide analogue of graphene, it was also
predicted that ZnO nanoribbons have ferromagnetic order in their ground
state.\cite{mendez1} In addition to these, it was reported that
common solvents can be used to exfoliate transition metal dichalcogenides and
oxides such as MoS$_2$, WS$_2$, MoSe$_2$, MoTe$_2$, TaSe$_2$, NbSe$_2$,
NiTe$_2$, Bi$_2$Te$_3$ and NbSe$_2$ to obtain single
layers.\cite{Coleman,mos2,ws2,pnas} Most recently, possibility of various
combinations of MX$_{2}$ type single-layer transition-metal oxides and
dichalcogenides, stable even in free-standing form, was
also predicted.\cite{mx2miz}

The recent synthesis of
silicene,\cite{silicene-exp1,silicene-exp2,silicene-exp3} the silicon analogue
of graphene has opened a new avenue to nanoscale material research. Though
the nanotube\cite{sint} and fullerene\cite{siful}forms of silicon were
synthesized earlier, monolayer silicon was presumed not to exist in a
freestanding form. Early theoretical works pointed out that silicene is a
semimetal with linearly crossing bands and it has a buckled crystal
structure that stems from $sp^{3}$ hybridization.\cite{seymur,hasan} Similar to
graphene, the hexagonal lattice symmetry of silicene exhibits a pair of
inequivalent valleys in the vicinity of the vertices of $K$ and $K^{\prime}$
symmetry points. Moreover, the experimental realization of the transformation of
thin films of wurtzite (WZ) materials into the graphene-like thin film structure
is another evidence for the existence of monolayer structures of Si and
Ge.\cite{stabilizing} Recent theoretical studies have revealed several
remarkable features of silicene such as a large spin-orbit gap at the  Dirac
point,\cite{silicene-soc} experimentally accessible quantum spin Hall
effect,\cite{silicene-qshe} transition from a topological insulating phase to a
band insulator that can be induced by an electric field\cite{silicene-ti}
and electrically tunable band gap.\cite{silicene-falko} In addition to unique
insulator phases such as quantum spin Hall, quantum anomalous Hall and band
insulator phases, the emergence of valley-polarized metal phase was also
reported very recently.\cite{motohiko} It appears that
silicene will be a possible graphene replacement not only due to its
graphene-like features but also because of its compatibility to existing
silicon-based electronic devices. 

In this paper, motivated by the very recent experimental realizations of
monolayer silicene,\cite{silicene-exp1,silicene-exp2,silicene-exp3} we
investigate how alkali, alkaline earth, and transition metal atoms interact
with monolayer freestanding silicene. Details of computational 
methodology are described in Sec. II. Characteristic properties of monolayer silicene and graphene are 
compared briefly. Our results on structural and electronic properties of metal adatom adsorbed silicene 
are presented in Sec. IV. Conclusions and a summary of our results are given in Sec. V.

\begin{figure}
\includegraphics[width=8.5cm]{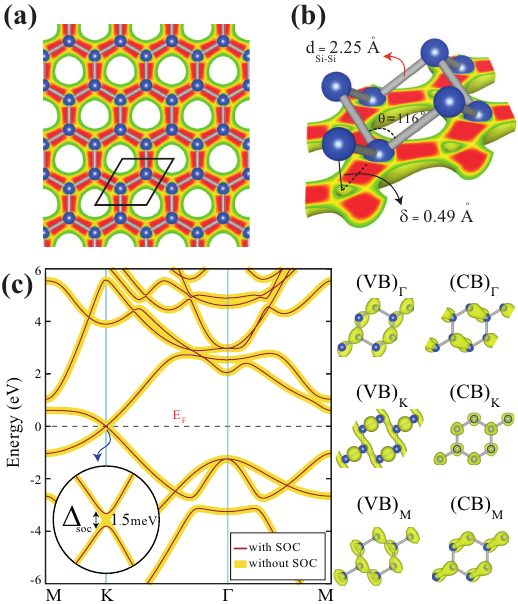}
\caption{(Color online) (a) Top view of the honeycomb lattice of silicene and
it's unitcell shown by black parallelogram. (b) Tilted view of 2x2 supercell of
silicene. Structural parameters: buckling ($\delta$), Si-Si-Si angle and Si-Si
distance are indicated. (c) Electronic band dispersion of silicene (with and
without SOC) and band decomposed charge densities of VB and CB at $\Gamma$, M
and K symmetry points.} \label{fig1}
\end{figure}

\section{Computational Methodology}

To investigate the adsorption characteristics of alkali metals and
transition metals on a monolayer honeycomb structure of silicone we employ
first-principles calculations\cite{kresse} using the projector augmented wave
(PAW) method\cite{paw} implemented in VASP code. Electronic exchange-correlation
effects are simulated using the spin polarized local density
approximation\cite{lda} (LDA). For the plane-wave basis set, the kinetic energy
cutoff is taken to be $ \hbar^2 | \mathbf{k}+\mathbf{G}|^2 / 2m = 500$ eV.
Brillouin zone (BZ) sampling is determined after extensive convergence analysis.
In the self-consistent potential, total energy and binding energy calculations
with a (6$\times$6$\times$1) supercell of silicene sheet a set of
(5$\times$5$\times$1) \textbf{k}-point
sampling is used for BZ integration. For partial occupancies the Gaussian
smearing method is used. The convergence criterion of our self consistent
calculations for ionic relaxations is $10^{-5}$ eV between two consecutive
steps. By using the conjugate gradient method, all atomic positions and the
size of the unitcell were optimized until the atomic forces were less than 0.05
eV/\AA. Pressures on the lattice unit cell are decreased to values less than 1
kB. Adatom adsorbed silicene monolayers are treated using a supercell geometry,
where a minimum of 15 \AA~ vacuum spacing is kept between the adjacent silicene 
layers. Diffusion pathway of adatoms are calculated for 10 different adsorption
points on (4$\times$4$\times$1) silicene supercell.

\begin{table}
\caption{Calculated values for graphene, and silicene. These are lattice
constant ($a$); Si-Si (or C-C) bond distance ($d$); thickness of the layer
($t$); workfunction ($\Phi$); cohesive energy per unit cell
($E_{coh}$); in-plane stiffness ($C$); optical phonon modes at the $\Gamma$
point.}
\label{table}
\begin{center}
\begin{tabular}{cccccccccccccccc}
\hline  \hline
    &$a$  & $d$& $\delta$ & $\theta$ & $\Phi$ & $E_{coh}$ & $C$& $Phonons$\\

Material & \AA  & \AA  & \AA & rad & $eV$ & $eV$ &$J/m^{2}$&$cm^{-1}$\\
\hline
Graphene & 2.46 & 1.42 &  -  & 120 & 4.49  &  17.87 &  335\cite{hasancf} &
900-1600\\
 \hline
Silicene & 3.83 & 2.25 & 0.49& 116 & 4.77 & 9.07  & 63\cite{strain}  &
150-580\cite{hasan} 
\\
\hline \hline
\end{tabular}
\end{center}
\end{table}

The cohesive energy of silicene (also for graphene) per unit cell 
relative to free Si atom, given in Table \ref{table}, is obtained from $E_{coh}
= 2E_T^{Si}-E_T^{Silicene}$, where $E_T^{Si}$ is the total energy of
single free Si and $E_T^{Silicene}$ is the total energy of silicene.
Here total energy of single atoms are calculated by considering their magnetic
ground state.  As for the adsorption energy of a metal
adatom, one can use the formula $E_{Ads} =
E_T^{Ad}+E_T^{Silicene}-E_T^{Silicene+Ad}$ where $E_T^{Silicene}$, $E_T^{Ad}$
and $E_T^{Silicene+Ad}$ are the total energies of (6$\times$6$\times$1)
supercell of silicene, isolated single adatom and silicene+adatom system,
respectively. 

Most of the
adatom adsorption result
in a net electrical-dipole moment perpendicular to the plane. Therefore ground
state electronic structure, magnetic state and workfunction are
calculated by applying a dipole correction\cite{dipole} to eliminate the
artificial electrostatic field between the periodic supercells. To obtain
continuous density of states curves and to determine the energy bandgap
($E_{g}$) smearing with 0.2 and 0.001 eV is used, respectively.  For the
charge transfer analysis, the effective charge on the atoms are obtained by the
Bader method.\cite{bader}

\section{Silicene}

Though crystalline silicon has the diamond structure and no layered form exists
in nature, very recent experimental studies have reported the successful
synthesis of a monolayer of silicon, called silicene, by the application of
various deposition techniques. Similar to graphene, silicene can be viewed as a
bipartite lattice composed of two inter-penetrating triangular sublattices of
silicon atoms. Since $\pi$ bonds between silicon atoms are weaker than in the
case of the carbon atoms, planarity is destabilized and therefore silicone atoms
are buckled in a silicene crystal. As shown in Fig. \ref{fig1}(b) the buckling
(perpendicular distance between these two Si planes) is 0.49 \AA. Upon the
formation of $sp^{3}$ bonded honeycomb lattice, the covalent bond length of
Si-Si is 2.25 \AA.

Two dimensional silicene sheet is a semimetal because the valence and conduction
bands touch at the Fermi level. It was predicted earlier that similar to
graphene, silicene has also linearly crossing bands at the $K$ (and
$K^{\prime}$) symmetry points and charge carriers in graphene behave like
relativistic particles with a conical energy spectrum with Fermi velocity
$V_{F}\cong10^{6} m/s$ like in graphene.\cite{seymur,hasan} In Fig.
\ref{fig1}(c), the electronic band structure of perfect silicene is presented.
Linear $\pi$ and $\pi^{*}$ bands that cross at the $K$ symmetry point are
responsible for the existence of massless Dirac fermions in silicene. Due to the
degeneracy of the valence band (VB) maxima and the conduction band (CB) minima
at the $K$ point, corresponding states have the same ionization potential and 
electron affinity. Therefore one can expect the observation of similar unique
properties of graphene in silicene. The calculated energy band gaps at $M$ and
$\Gamma$ symmetry points are 1.64 and 3.29 eV, respectively. At the $\Gamma$
point, the degenerate VB is composed of $p_{x}$ and  $p_{y}$ orbitals, while the
CB is formed by the hybridization of $s$ and $p_{z}$ orbitals. 

\begin{figure}
\includegraphics[width=5cm]{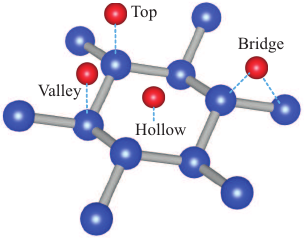}
\caption{(Color online) Preferable adsorption sites: Hollow, Top, 
Hill and Bridge on a silicene lattice.}  \label{fig2}
\end{figure}

In Fig.\ref{fig1}(c) we also present the electronic band dispersion taking into
account spin-orbit coupling (SOC). Though clearly the inclusion of
spin-orbit interaction does not result in a visible change in band dispersion, a
band gap of $\Delta = 1.5$ meV appears at the K point. Due to its buckled
structure, silicene has a larger SO-induced gap than graphene which is of the
order of 10$^{-3}$ meV.\cite{soc-gr} The calculated band structure and band gap
opening ($\Delta$) with SOC is in good agreement with recently reported
studies.\cite{silicene-qshe, silicene-soc, motohiko}

In Table \ref{table} we also compare structural, electronic and vibrational
properties of graphene and silicene. It appears that in contrast to the general
trend that the larger the atomic radius, the smaller the workfunction,
silicene's workfunction is 4.77 eV, while for graphene it is 4.49 eV. The
calculated values of in plane stiffness\cite{strain}  and cohesive energy
indicates that silicene is a stable but less stiffer material as compared to
graphene. Similar to graphene's out-of-plane optical (ZO) phonon mode at 900
$cm^{-1}$, silicene has a ZO mode at 150 $cm^{-1}$. Eigenfrequency of the LO and
TO modes are degenerate at the $\Gamma$ symmetry point are found to be 580
$cm^{-1}$ which is almost three times smaller than graphene's LO (and TO) modes.

\begin{figure}
\includegraphics[width=8.5cm]{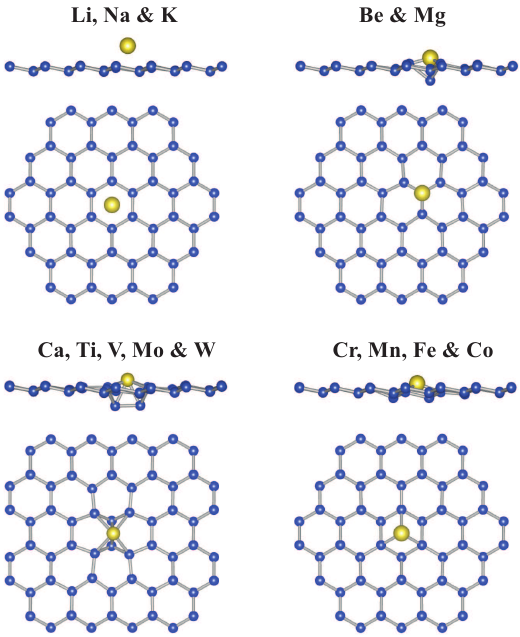}
\caption{(Color online) Side and top view for characteristic adsorption 
geometries for alkali, alkaline earth and transition metal atoms.} 
\label{fig3}
\end{figure}

\section{Results: Adsorption of Metal Atoms}

As shown in Fig.~\ref{fig2}, regarding the interaction of silicene surface with
adsorbates four different adsorption positions can be considered i.e., above the
center of hexagonal silicon rings (Hollow site), on top of the upper silicon
atoms (Top-site), on top of the lower silicon atoms (Valley-site), on top of the
Si-Si bond (Bridge site). Considering the monolayer hexagonal lattice structure
of silicene, it is reasonable to expect the relaxation of foreign atoms
to one of these adsorption sites.

\subsection{Bonding Geometry and Migration Barriers}

\begin{figure}
\includegraphics[width=8.5cm]{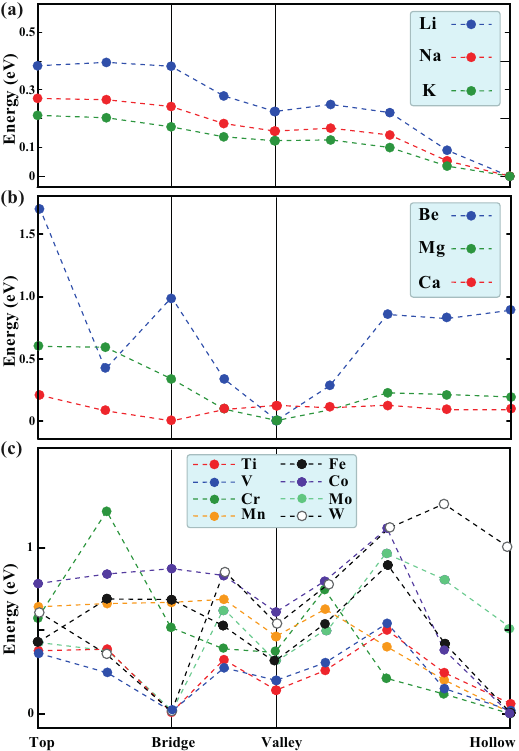}
\caption{(Color online) Diffusion pathways and barriers of (a) alkali atoms and
(b) $3d$, $4d$ and $5d$ transition metals through top, bridge, valley and hollow
sites.} 
\label{fig4}
\end{figure} 

We first investigate the adsorption characteristics of alkali adatoms Li, Na and
K on silicene. The alkali metals are highly reactive metals and their chemical
activity increases from Li to Fr. Characteristic bonding geometry of alkali
atoms are depicted in Fig.\ref{fig3}. Upon full geometry 
optimization, all alkali atoms Li, Na and K atoms favors bonding on the hollow
site of the silicene layer. Adsorption of alkali atoms does not yield any
significant distortion or stress on the silicene lattice. The valley site on the
low-lying silicon atoms is the next favorable site. Though the top and
valley site adsorptions are also possible, bridge site adsorption of alkalis is
not possible on a silicene 
lattice. Therefore the bridge site adsorption is a kind of transition state
between top and valley sites. Structural and electronic properties of alkali
metals adsorbed on silicene layer are also presented in Table \ref{table2}. Here
the height of adatoms is calculated as the difference between the 
average coordinates of neighboring Si atoms and the adsorbate. The distance
between the adatom and silicene surface monotonically increases with increasing
atomic size.
However, fully conforming to graphene,\cite{chan} there is no clear trend in the
adsorption energies. While adsorption energies of Li, Na and K on graphene were
calculated to be 1.1, 0.5 and 0.8 eV, respectively, their binding to silicene
lattice is more than twice stronger, i.e. 2.4, 1.9 and 2.1 eV, respectively. The
nature of the alkali-silicene bond will be discussed in the next subsection.

Possible diffusion pathways of the adsorbate atoms on silicene lattice can 
also be deduced from the energy barrier between hollow, valley, top and bridge
sites. In Fig. \ref{fig4} we present the energetics of different
adsorption sites. For alkali atoms, shown in Fig. \ref{fig4}(a) it appears that
the most likely migration path between subsequent hollow sites passes thorough
the nearest valley sites.  It is also seen that the energy difference between
hollow and valley sites becomes smaller for larger atoms and therefore the
diffusion of larger alkali atoms are relatively easier. Though alkali atoms
strongly bind to the silicene surface, at high temperatures they may diffuse
along hollow and valley sites when they overcome the energy barrier of 140-280
meV.

Alkaline earth metals are the elements of the periodic table having two valence
electrons in their outermost orbital. Compared to alkalis, alkaline earth metals
have smaller atomic size, higher melting point, higher ionization energy and
larger effective charge. Strong interaction between alkaline earths and silicon
surfaces is well-known and have been used for various silicon etching and
surface engineering techniques. Therefore one can expect strong bonding of
alkaline earth's to a monolayer silicon surface. Resulting atomic geometries for
Be, Mg and Ca adsorbed on a silicene sheet are depicted in Fig. \ref{fig3}.
Unlike alkalis, the hollow site is not the most favorable adsorption site for
alkaline earth metals. Among these, while Be and Mg favor adsorption to a valley
site, the Ca adsorbate that has a quite large atomic size (empirically $\sim$
1.8 \AA) prefers bridge site adsorption. Adsorption energy of alkaline earth's
is slightly higher for alkali metals (except Mg). Similar to the Na adsorbate a
second row element, adsorption energy has a sharp decrease for the second row
alkali element Mg. As shown in Fig. \ref{fig4}(b), the Be
atom has to overcome a quite large energy barrier ($\Delta$E > 1 eV) for
migration from valley to other adsorption sites. Sudden increases in migration
barrier stem from the stretching of silicene lattice by adsorbate which is
freed only in one direction perpendicular to the surface. At high
temperatures, migration of Ca atoms through bridge and hollow sites may take
place by overcoming the energy barrier of 125 meV.

\begin{table}
\caption{Calculated values for adatom adsorption on (6$\times$6$\times$1)
silicene; relaxation
sites hollow (H), bridge (B), valley (V) or top (T), adatom height ($h$),
adsorption energy of adatom ($E_{Ads}$), total magnetic moment of the system
($\mu_{tot}$) in units of Bohr magneton ($\mu_{B}$), energy band gap ($E_{g}$),
dipole moment (\textbf{p}), Bader charge transfered from adatom to silicene
($\rho_{ad}$) and the workfunction of the optimized structure
($\Phi$). Metallic and half-metallic structures are denoted as \textbf{m} and
\textbf{hm}, respectively.} \label{table2}
\begin{tabular}{ccccccccccc}
\hline  \hline
 &Site & $h$ &$E_{Ads}$ & $\mu_{iso}$& $\mu_{tot}$&$E_{g} $ &
\textbf{p}&$\rho_{ad}$ & $\Phi$\\
 & & (\AA) & ($eV$) & $(\mu_{B})$ & $(\mu_{B})$ & ($eV$)& ($e$\AA) & ($e$) &
($eV$)\\

\hline
\hline
Li  & \textbf{H} &  1.69  & 2.40  & 1.0  &  0.0  & \textbf{m}  & 0.30  & 0.8 & 
4.39   \\
Na  & \textbf{H} &  2.19  & 1.85  & 1.0  &  0.0  & \textbf{m}  & 0.60  & 0.8 & 
4.25   \\
K   & \textbf{H} &  2.70  & 2.11  & 1.0  &  0.0  & \textbf{m}  & 0.94  & 0.8 & 
4.09   \\
\hline
Be  & \textbf{V} &  0.78  & 2.87 & 0.0 & 0.0 & 0.39  & 0.00  & 1.3 &4.68  \\
Mg  & \textbf{V} &  1.98  & 1.22 & 0.0 & 0.0 & 0.48  & 0.31  & 1.0 &4.81  \\
Ca  & \textbf{B} &  1.49  & 2.68 & 0.0 & 0.0 & 0.17  &0.62& 1.3 &4.47  \\
\hline
Ti  & \textbf{B} & 0.77 & 4.89 & 4.0 & 2.0 & \textbf{hm} & 0.29 & 0.9 &4.77\\
V   & \textbf{B} & 0.62 & 4.32 & 5.0 & 2.7 & 0.06 &  0.18 & 0.7  & 4.84  \\
Cr  & \textbf{H} & 0.48 & 3.20 & 6.0 & 4.0 & \textbf{hm}& 0.12  & 0.3 &4.65\\
Mn  & \textbf{H} & 1.04 & 3.48 & 5.0 & 3.0 & 0.24  & 0.10 & 0.4 &  4.73  \\

Fe  & \textbf{H}& 0.33 & 4.79 & 4.0 & 2.0 & 0.18  & 0.00 & 0.0 & 4.81  \\
Co  & \textbf{H}& 0.72 & 5.61 & 3.0 & 1.0 &\textbf{m}& 0.00 & 0.0 & 5.00  \\
Mo  & \textbf{B}& 0.37 & 5.46 & 6.0 & 0.0 &\textbf{m}& 0.15 & 0.1 & 4.97\\
W   & \textbf{B}& 0.01 & 7.05 & 6.0 & 0.0 &  0.02 & 0.04 & 0.2 & 4.90  \\
\hline
\hline

\end{tabular}
\end{table}

We next investigate the adsorption characteristics of eight elements of the
$3d$, $4d$ and $5d$ transition metal adatoms: Ti, V, Cr, Mn, Fe, Co, Mo and W.
Though the outermost $s$ orbitals of the transition metals are completely
filled, because of their partially filled inner $d$ orbitals diverse adsorption
characteristics for different atoms can be expected. Due to relatively small
atomic radius of all transition metals and having more electrons that can
participate in the chemical bonding, we can expect stronger binding to the
silicene lattice. From Fig. \ref{fig3} it appears that while the adsorption of
alkalis do not cause any significant change in the silicene lattice, transition
metals are more likely to disturb the nearest silicon atoms.

Bonding of Ti adatom with 4.89 eV to silicene occurs with a significant lattice
distortion whereas the adsorption of Ti occurs on the hollow site of graphene
without disturbing the planar lattice structure. As a consequence of bridge-site
adsorption of Ti, it binds six nearest Si atoms strongly by pushing the
underlying two Si atoms downwards. In order not to exclude possible vacancy
formation and adatom-induced fracturing in such Ti-adsorbed silicene lattice we
also examine the stability of whole structure through molecular
dynamics (MD) calculations. Ab initio MD calculations show that the adsorbed Ti
atom remains bounded and neighboring Si-Si bonds are not broken after $\sim$1
$ps$ at 300 K. Similar to Ti, V adatom is adsorbed on bridge site with 4.32 eV
adsorption energy.

Among the $3d$ transition metal adatoms considered here only Cr, Mn, Fe and Co
similar to alkalis, are adsorbed on the hollow site. However, differing from
alkalis, transition metals have quite strong binding (3.20, 3.48, 4.79 and 5.61
for Cr, Mn, Fe and Co, respectively) with three uppermost Si atoms. Therefore,
instead of being adsorbed on top a hollow site like alkalis, transition metals
are almost confined in the silicene plane. Energetics of transition metal
atoms on the most favorable adsorption sites are also listed in Table
\ref{table2}.

Note that not only the electronic occupancy but also the atomic radius of the
atom is important in determining the final geometry of the adatom on silicene.
Since for $d$ orbitals having more than half-occupancy the atomic radii are
significantly decreased by increasing the number of electrons, starting from Cr
all 3$d$ transition metals prefer to be relaxed on hollow site. To see how the
atomic radius of transition metal affects the adsorption geometry we also
perform calculations for other Group 6 elements Mo and W. According to the most
recent measurement of Cordero \textit{et al.}\cite{radius} covalent atomic radii
of Cr, Mn and W are 1.39, 1.54 and 1.62 \AA, respectively. In this column of the
Periodic Table, while Cr is adsorbed on hollow site, bridge site adsorption
becomes more preferable for Mo and W due to their larger atomic radii. Effect of
the atomic radii can also be seen even in the same row elements: adatoms Ti and
V, that have covalent radius larger than that of Cr, are relaxed to the bridge
site. Thus we can infer that only the transition metal adatoms having covalent
atomic radii larger than $\sim$1.50 \AA~favor bridge site adsorption.

It appears from Fig. \ref{fig4}(c) that the energy barrier between different
adsorption sites is relatively large for transition metal atoms. The diffusion
barrier between the most favorable site and the less favorable one is $\sim$0.6
eV. The most likely diffusion path for Ti, V, Mo and W passes from bridge to
hollow sites. However, diffusion of Mn, Cr, Fe and Co atoms from one hollow site
to another one may occur via valley sites.

Furthermore, the effect of spin-orbit interaction on the optimized 
adsorbate-silicene geometry is examined for adsorption on (6$\times$6$\times$1)
silicene supercell. Compared with the optimized geometries obtained excluding
SOC, SO-induced change in adatom-silicon bond length is just on the order of
0.001 \AA. It is also found that the migration path profile and the most
favorable adsorption site do not change when SOC is included. Therefore, due to
negligible effect of spin-orbit interaction on structural properties, in the
rest of our study LDA calculations will be employed.

\begin{figure}
\includegraphics[width=8.5cm]{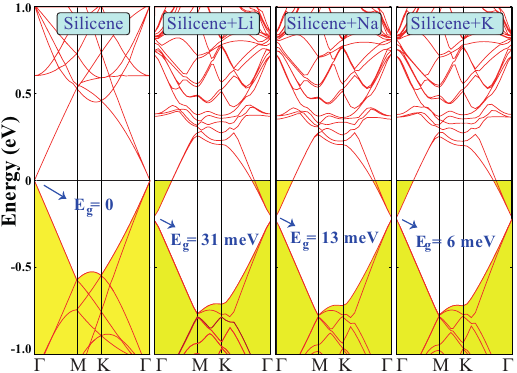}
\caption{(Color online) Electronic band structures for perfect, Li, Na and 
K adsorbed silicene. Fermi level is set to zero. Occupied bands are filled with
yellow color.}  \label{fig5}
\end{figure}

\subsection{Electronic Structure}

In this section we present spin-polarized electronic band dispersion and density
of states (DOS) for adatoms adsorbed on the most favorable site on the silicene
surface.  Since the alkaline earth and transition metal adsorption
significantly disturb the hexagonal lattice symmetry, electronic band dispersion
along the high symmetry points ($\Gamma$-M-K-$\Gamma$) of perfect silicene may
not represent the real electronic properties of the whole structure. Therefore,
a density of states plot covering large number of k-points in the BZ is more
convenient for a reliable description of the electronic structure.

In Fig. \ref{fig5}, we present electronic band dispersions of alkali
metal adsorbed silicene. For the sake of comparison, the band structure of the
bare silicene is also included. Here it is worth to note that the Dirac cone
that is located at 2/3 of first BZ is shifted from $K$ to $\Gamma$ high symmetry
point due to band folding of the (6$\times$6$\times$1) silicene supercell.
Similar shift can
be obtained by band folding operations for all (3n$\times$3n$\times$1)
supercells ($n$=1,
2,...). As a result of the adsorption of an alkali atom, semimetallic silicene
becomes metallic due to the donation of $\sim$0.8 e charge from the alkali atom
into the silicene conduction band. Such an attraction of adatom through the
hollow site of the silicene surface with a significant charge transfer resembles
ionic bonding. Moreover, since the band dispersion of perfect silicene is
negligibly disturbed by the alkali adatoms, it seems reasonable to assume that
the bonding is of ionic type. Here, in accordance to the charge transfer, the
Fermi level (E$_{F}$) is shifted by $\sim$0.2 eV. It is seen that all alkali
atom adsorption bands formed by the hybridization of adsorbate-$s$ states with
the silicene-$p$ states appear in the vicinity of 0.4 eV. Additionally, a small
gap opening in the $p$ bands below the Fermi level, which is 6, 13 and 31 meV
for Li, Na and K, respectively, occur at the crossing point.

While each isolated alkali atom has a net initial magnetic moment of 1
$\mu_{B}$, up-down spin degeneracy is not broken upon 0.8 $e$ charge transfer
and therefore all the silicene+alkali structures are nonmagnetic metals.
However, as a result of the large charge transfer between the alkali atom and
silicene, remarkable dipole moment perpendicular to the silicene surface is
induced. Calculated value of the dipole moment directed from silicene to the
adatom is 0.30, 0.60 and 0.94 e$\cdot$\AA~for Li, Na and K, respectively. Though
the amount of the charge transfered is almost the same for all alkalis, the
induced
dipole moment is different due to the difference in distance between the
adsorbate and the silicene layer. As a consequence of adatom-induced surface
charge density redistribution, the workfunction of the silicene surface is
decreased. Therefore, the trend of the increase in the dipole moment follows the
decrease in workfunction.

\begin{figure}
\includegraphics[width=8.5cm]{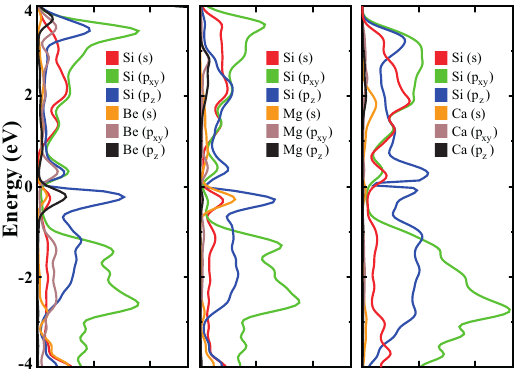}
\caption{(Color online) Partial DOS for Be, Mg and Ca adsorbates and nearest
silicon atoms. Fermi level is set to zero. DOS is broadened by Gaussian smearing
with 0.2 eV.}  \label{fig6}
\end{figure}

Alkaline earth metals are highly reactive elements that tend to form various 
compounds by loosing two outermost shell electrons. Natural silicate and
carbonate forms of alkaline earths are well-known. As shown in Table
\ref{table2}, Be, Mg and Ca adsorption yields a gap opening of 0.39, 0.48 and
0.17 eV, respectively. It is also
seen from the partial density of states  shown in Fig. \ref{fig6} that the top
of the valence band (VB) of Be adsorbed silicene is due to the hybridization of
$s$ and $p_{z}$ orbitals of the nearest silicon atoms and the alkaline earth
metals. However, at top of the VB main contribution comes from Mg-$s$, Si-$s$
and Si-$p_{z}$ hybridization, whereas $p_{z}$ orbital of Mg does not mix with
surrounding silicon states. The bottom of the conduction band (CB) of both Be
and Mg is formed by hybridization of $p_{xy}$ states of adsorbates with $p_{xy}$
and $p_{z}$ states of silicene. Differing from Be and Mg, adsorption of Ca on
silicene does not yield appearance of Ca states around E$_{F}$. While the states
at the VB edge arise mainly from the silicene $p_{z}$ orbitals, the CB edge is
composed of $s$, $p_{xy}$ and $p_{z}$ orbitals of Si.

\begin{figure}
\includegraphics[width=8.5cm]{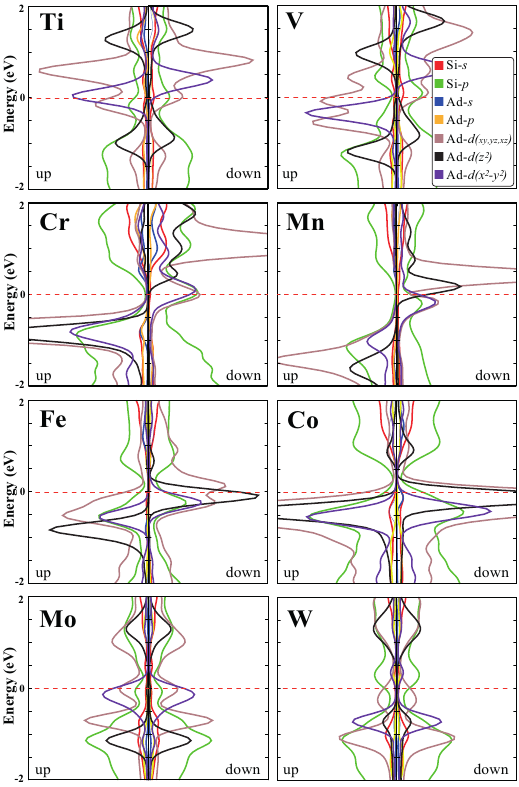}
\caption{(Color online) Partial DOS for Ti, V, Cr and Mn adsorbates and 
nearest silicon atoms. Fermi level is set to zero. DOS is broadened by Gaussian
smearing with 0.2 eV.}  \label{fig7}
\end{figure}

Since atomic $d$ states of transition metal atoms have comparable energy values
to that of their $s$ valence states, the $d$ shell of these atoms is partially
filled. The net magnetic moment of isolated transition metals are nonzero and
behave like small magnets unless the $d$ shell is completely filled. Though the
$d$ shell electrons are located close to the nucleus like the characteristic
core electrons, they can spread out much further like valence electrons.
Therefore, transition metals with their partially filled $d$ shells are relaxed
to different sites on silicene and we can expect diverse electronic properties
upon their adsorption. We depict the spin-polarized electronic density of states
of those transition metal adsorbed silicene in Fig \ref{fig7}.

Ti adsorption on silicene has a quite different characteristics as 
compared to other transition metals. Upon the adsorption of a Ti atom that has
an initial magnetic moment of 4.0 $\mu_{B}$ in its isolated state, on bridge
site of silicene the structure becomes a ferromagnetic metal with 2.0 $\mu_{B}$
net magnetic moment per supercell. Since we use a quite large supercell such
that the adatom-adatom interactions are negligible, describing the adatom levels
in terms of orbitals, instead of bands, is more appropriate. As a result of the
strong bonding, the silicene lattice is significantly distorted, Ti atom binds
to six nearest Si atoms and the degeneracies of the Ti-$d$ states are completely
broken. Metallic bands of the structure originates only from the
Ti-$d_{x^{2}-y^{2}}$($\uparrow$) orbitals and silicene becomes half metal 
when Ti adatom is adsorbed.

For the case of V adsorption on silicene, the structure becomes 
semiconductor with 2.7 $\mu_{B}$ net magnetic moment per supercell. It is seen
from Fig \ref{fig7} that while only V-$d_{xy,yz,xz}$($\uparrow$) states
have a significant contribution to VB maximum, bottom of the CB is derived from
Si-$p$($\downarrow$) and V-$d_{xy,yz,xz}$($\downarrow$). The contribution of
the other eigenstates at the vicinity of E$_{F}$ are enhanced just because of
the smearing of the DOS plot. Cr adsorption also turns the
semimetallic silicene into a half metallic material with 4 $\mu_{B}$ per
supercell. The up spin (minority) bands of the Cr adsorbed silicene are
semiconducting with a direct band gap of $\sim$0.4 eV, whereas the down spin
(majority) band shows metallic
behavior. The bands in the vicinity of E$_{F}$ are mainly derived 
from the hybridization of Cr-$d_{xy,yz,xz}$($\downarrow$) and
Cr-$d_{x^{2}-y^{2}}$($\downarrow$) orbitals and the Si-$p$($\downarrow$)
states. However, differing from Cr, the adsorption of Mo and W that occurs on
bridge site results in a nonmagnetic ground state. While Mo-adsorbed silicene
becomes metallic due to Mo-$d_{xy,yz,xz}$ states at E$_{F}$, W-adsorbed
silicene is a nonmagnetic semiconductor with bandgap of 0.02 eV. Note that the
half-metallic behavior for Ti- and Cr-decorated silicene can be quite important
for potential use in spintronics.

The adsorption of Mn atom in close proximity to the hollow site of buckled
hexagonal lattice results in semiconducting behavior with a bandgap of 0.24 eV.
The degeneracy of $\uparrow$ and $\downarrow$ spin states are broken due to the
existence of a net magnetic moment of 3.0 $\mu_{B}$. Since Mn, Fe and Co atoms
have the smallest atomic size of the considered metal atoms here, they are the
most closely bonded one among all hollow site adsorbed atoms. Both the
adsorption of Fe and Co result in metallic ground state, that originates from
the $d_{xy,yz,xz}$($\downarrow$) and $d_{z^{2}}$($\downarrow$) states at
E$_{F}$, with 2 and 1 $\mu_{B}$ net magnetic moment, respectively. As a
consequence of the small adsorption height, the induced dipole moments are much
smaller than those of the alkali adatom adsorbed silicene.

The adsorbate-induced magnetic moments induced in the silicene layer lead to an
exchange-splitting especially in $d$ eigenstates. The calculated value of the
exchange-splitting is 0.4, 2.0, 2.1, 1.9, 0.5 and 0.4 eV for Ti, V, Cr, Mn, Fe
and Co, respectively. However, bridge-site adsorbed Mo and W adatoms exhibits no
splitting due to their nonmagnetic ground state.

\begin{figure}
\includegraphics[width=8.5cm]{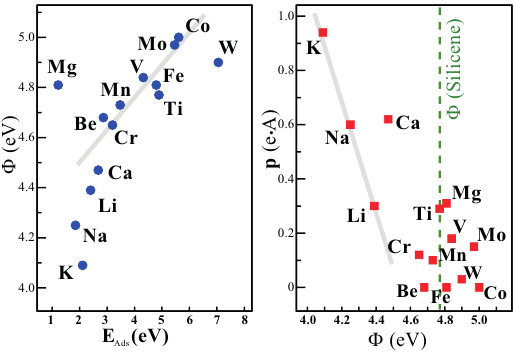}
\caption{(Color online) (a) Plot of the adsorption energy E$_{Ads}$ versus 
the workfunction $\Phi$ and (b) the workfunction $\Phi$ versus dipole moment
\textbf{p}.}  \label{fig8}
\end{figure}

To reveal the correlations between adsorption energy (E$_{Ads}$), 
workfunction ($\Phi$) and induced dipole moment (\textbf{p}) we also present
$\Phi$-E$_{Ads}$ and \textbf{p}-$\Phi$ plots as shown in  Fig \ref{fig8}. Note
that the adsorption of large transition metals that occurs with large binding
energy changes the silicene's workfunction negligibly. Especially for $3d$
transition metals having more than half filled $d$-subshell, the larger the
adsorption energy the higher the workfunction. However, it appears that even
with a small coverage (1/72) of silicene surface by alkali metals, a significant
decrease of the work function can be produced. For alkalis and alkaline earths,
there is almost a linear correlation between E$_{Ads}$ and $\Phi$. Because of
the ionic nature of the alkali-silicene bonding, the workfunction linearly
depends on the atomic size and therefore one can expect a significant decrease
(>1 eV) in the workfunction for adsorption of larger alkalis on silicene. It is
also seen that, while multivalent adatoms do not follow a particular trend in
dipole moment, for alkali atoms workfunction shift exhibits a linear decrease
with increasing dipole moment.

\section{Conclusions}

In this study, we presented a first-principles investigation of the 
adsorption characteristics of alkali (Li, Na and K), alkaline earth (Be, Mg and
Ca) and transition metals (Ti, V, Cr, Mn, Fe, Co, Mo and W) on silicene.
Investigation of the interaction of silicene with metal atoms has a significant
importance because of its fundamental relevance to applications in catalysis,
batteries and nanoelectronics.

We found that silicene has quite different adsorption characteristics as 
compared to graphene\cite{weh1,weh2} because of the diversity of adsorption
geometries and its
high surface reactivity. As a consequence of its buckled lattice structure, all
the metal atoms bind strongly to silicene. While the binding energy of alkali
and alkaline earth metals is 1-3 eV, transition metals have larger binding
energies of 3-7 eV. Our diffusion path analysis shows that compared to graphene
migration of metal atoms on a silicene lattice is more difficult and requires to
overcome higher energy barriers. On the other hand, diffusion of metal atoms Li,
Na, K, Mg and Ca may occur at moderate temperatures. Depending on the adatom
type, semimetallic silicene can show metal, half-metal and semiconducting
behavior. It was also noted that as a result of the high surface reactivity the
direction of the charge transfer is always from metal adsorbate to silicene.
However, the existence of charge donation and the resulting adatom-induced
dipole modify the workfunction of silicene considerably. Especially the linear
decrease in $\Phi$ for larger alkali adatoms is promising for tunable
enhancement of electron and ion emission which is appealing to silicene-based
thermionic converters and cathodes. Our findings also suggest that the
half-metallic ferromagnetic nature of Ti- and Cr-decorated silicene have a great
potential for silicon-based spintronic device applications. Moreover the
existence of a tunable bandgap opening in silicene by alkaline earth
metal adatoms is highly desirable for its use in nanoscale optoelectronic device
applications.

The extension of our investigation to a detailed study to include the effect of
spin-orbit coupling and of the substrate interactions on the adsorption
characteristics of silicene is planned for future studies.

\section{Acknowledgements}

This work was supported by the Flemish Science Foundation (FWO-Vl). 
Computational resources were provided by TUBITAK ULAKBIM, High Performance and
Grid Computing Center (TR-Grid e-Infrastructure). H. S. is supported by a FWO
Pegasus Marie Curie Fellowship.

\end{document}